\documentclass[aps,twocolumn,nofootinbib,showpacs]{revtex4}

\usepackage{latexsym}
\usepackage{graphicx}
\usepackage{subfigure}
\usepackage{hyperref}

\usepackage{color}% use if color is used in text

\begin{document}

\title{Radio bursts from superconducting strings}

\author{Yi-Fu Cai}
\email{ycai21@asu.edu}
\author{Eray Sabancilar}
\email{Eray.Sabancilar@asu.edu}
\author{Tanmay Vachaspati}
\email{tvachasp@asu.edu}

\affiliation{
Physics Department, Arizona State University, Tempe, Arizona 85287, USA.
}

\begin{abstract}

We show that radio bursts from cusps on superconducting strings
are linearly polarized, thus, providing a signature that can be used
to distinguish them from astrophysical sources. We write the event 
rate of string-generated radio transients in terms of observational 
variables, namely, the event duration and flux. Assuming a canonical 
set of observational parameters, we find that the burst event rate 
can be quite reasonable, e.g., order ten a year for Grand Unified 
strings with 100 TeV currents, and a lack of observed radio bursts can
potentially place strong constraints on particle physics models.
\end{abstract}

\maketitle

%\section{Introduction}

Cosmic strings are possible relics from the early Universe. Their
discovery would substantiate our hot big bang cosmological model
and also provide tremendous insight into the nature of fundamental
interactions.

There are a number of different ways to look for cosmic strings,
mostly based on their gravitational interactions, and negative searches
so far impose constraints on particle physics models and cosmology.
If the strings are superconducting \cite{Witten:1984eb}, their
electromagnetic emission provides yet another signature that can
be used to search for them. The electromagnetic emission from a
cosmic string loop is not steady and can have sharp bursts that
can be seen as transient events. In Ref.~\cite{Vachaspati:2008su},
it was pointed out that it might be fruitful to look for superconducting
strings by searching for bursts at radio wavelengths. There is a simple
reason for choosing to look in the radio band. Cosmic strings are
large objects and their fundamental frequency of emission is very
low. The power emitted at higher frequencies generally falls off with 
increasing harmonic. Thus, there is more power emitted in the radio than 
in other bands such as
the optical. Also, as we shall see in Sec.~\ref{burst},
the emission in the burst is beamed, with the beam being widest at
lower frequencies. Thus, the event rate in radio bursts can be expected
to be larger than those at higher frequencies. On the other
hand, propagation effects in the radio band are stronger, and these
have to be included when evaluating the signature. 

Besides superconducting cosmic strings, there are other strong
motivations for looking at transient radio phenomenon from pulsars,
supernovae, black hole evaporation, gamma-ray bursts, active galactic
nuclei, and extra-terrestrial life. A radio burst from a
superconducting cosmic string will have to be distinguished among
bursts from other potential astrophysical sources. With this in mind,
we recalculate the characteristics of the string burst and show that
it is linearly polarized in a direction that is independent of
the frequency.

The feasibility of observations depends on the event rates for
radio bursts. Here, we focus on evaluating the event rate in variables
that are most useful to observers. The burst at source occurs
with a certain duration and flux. However, the observed duration 
and flux depend on the redshift of the source. We transform the event 
rate from source variables to observer variables. 
These results will be useful in the ongoing search for radio transients
at the Parkes \cite{Lorimer:2007qn}, ETA \cite{eta}, and LWA \cite{lwa} 
telescopes, and the new generation large radio telescopes such as 
LOFAR\cite{lofar} and SKA \cite{ska}.

This paper is organized as follows. In Sec.~\ref{burst}, we calculate
the characteristics of a burst from a superconducting string,
including the polarization. In Sec.~\ref{sec:eventrate}, we find the
event rate in observer variables, followed by a numerical evaluation
in Sec.~\ref{sec:numerical}. We conclude in Sec.~\ref{conclusions}.

\section{Burst characteristics}
\label{burst}

The electromagnetic field due to a superconducting cosmic string
is given by Maxwell's equations
\begin{equation}
\partial_\nu\partial^\nu A^\mu = 4\pi J^\mu,
\end{equation}
where
\begin{equation}
J^\mu(t,{\vec x}) = I \int d\sigma f^\mu_{,\sigma}
                         \delta^{(3)}({\vec x}-{\vec f} (t,\sigma)),
\label{jmu}
\end{equation}
and we have assumed that the string denoted by $f^\mu(t,\sigma)$
carries a uniform and constant current $I$.

A string loop oscillates and the general solution in its center of
mass frame can be written as
\begin{eqnarray}
f^0&=&t, \ \
{\vec f}(t,\sigma) = \frac{1}{2}\left [
      {\vec f}_+ (\sigma_+) + {\vec f}_-(\sigma_-) \right ],
\nonumber \\
\sigma_\pm &=& \sigma \pm t, \ \ | {\vec f}_{\pm}' | = 1 , \ \
\int d\sigma_\pm {\vec f}_\pm = 0,
\label{stringmotion}
\end{eqnarray}
where a prime denotes derivate with respect to the argument.

The power emitted in electromagnetic radiation from superconducting
strings has been analyzed in \cite{Vilenkin:1986zz} but the
polarization has not been studied. So, we repeat earlier
analyses to show that the radiation is linearly polarized. This also
leads into the analysis of the event rate in Sec.~\ref{sec:eventrate}.

The string dynamics is periodic and so is the current. Hence we work
with discrete Fourier transforms
\begin{eqnarray}
 A^{\mu}(t,\vec{x}) &=&
\sum_\omega e^{-i\omega t}A_\omega^{\mu}(\vec{x})~,\\
 J^{\mu}(t,\vec{x}) &=&
\sum_\omega \int d^3k e^{-i(\omega t-\vec{k}\cdot\vec{x})}
J_\omega^{\mu}(\vec{k})~,
\end{eqnarray}
where $\omega=4\pi n/L$ and $n$ is an integer. Then
\begin{eqnarray}
A_\omega^\mu(\vec{x}) =
\int d^3x' \int d^3k
\frac{e^{i\vec{k}\cdot\vec{x}'}}{|\vec{x}-\vec{x}'|}
J_\omega^\mu(\vec{k})e^{i\omega|\vec{x}-\vec{x}'|}~,
\label{eq:solution}
\end{eqnarray}
and $J_\omega^\mu ({\vec k})$ follows from Eq. (\ref{jmu})
\begin{eqnarray}\label{Jwmu}
 J_\omega^\mu(\vec{k}) =
\frac{2I}{(2\pi)^3L}
\int_0^{L/2}dt \int_0^Ld\sigma
e^{i(\omega t-\vec{k}\cdot\vec{f})}f'^{\mu}(t,\sigma)~,
\end{eqnarray}
where the delta function that appeared in Eq. (\ref{jmu}) has been integrated
out. In terms of the left and right movers of Eq.~(\ref{stringmotion}),
we get
\begin{eqnarray}
 J_\omega^\mu(\vec{k}) =
\frac{2I}{(2\pi)^3L} ({J_+}^\mu{J_-}^0 + {J_+}^0{J_-}^\mu ),
\end{eqnarray}
where
\begin{eqnarray}
 J_\pm^\mu(\vec{k}) =
\int_0^{L} d\sigma_\pm e^{ik\cdot f_\pm/2} f_\pm'^{\mu}.
\label{eq:Jk}
\end{eqnarray}

The loops that will give us the strongest observational signatures
will have lengths that are much larger than the radio wavelengths
at which observations are made. Hence $\omega L = 4\pi n \gg 1$,
and only high harmonics are of interest. The integrals $J_\pm$ can
be evaluated using the saddle point approximation. If the saddle
point has a nonvanishing imaginary piece in the complex $\sigma_\pm$
plane, the integrals fall off exponentially fast with $n$, and the
electromagnetic radiation in those directions, ${\vec k}$, is
negligible. The interesting situation is when the saddle point
is real in the evaluation of both $J_\pm^\mu$. This can happen if
\begin{equation}
k \cdot f'_\pm = 0,
\end{equation}
and corresponds to a ``cusp'' on the string loop as discussed in
earlier work, and more recently in some generality in
Ref.~\cite{Steer:2010jk}. The integrals can be evaluated by expansion
around the real saddle point and lead to
\begin{eqnarray}\label{Jwmue}
 J_\omega^\mu(\vec{k}) \simeq i\frac{2I}{(2\pi)^3\omega}e_\omega^\mu~,
\end{eqnarray}
where
\begin{eqnarray}
 e_\omega^\mu = -\frac{i}{L\alpha_+\alpha_-} \bigg[ \frac{f''^\mu_+}{\alpha_+}
\int du_+u_+e^{iu_+^3} \int du_-e^{-iu_-^3} + \nonumber\\
 \frac{f''^\mu_-}{\alpha_-} \int du_+e^{iu_+^3} \int du_-u_-e^{-iu_-^3} \bigg]~,
\end{eqnarray}
and
\begin{eqnarray}
 u_+=\omega^{1/3}\alpha_+\sigma_+~,~~\alpha_+=(\frac{l_\mu f'''^\mu_+}{12})^{1/3}~,
\nonumber\\
 u_-=\omega^{1/3}\alpha_-\sigma_-~,~~\alpha_-=(\frac{l_\mu f'''^\mu_-}{12})^{1/3}~,
\end{eqnarray}
where $k\equiv \omega l^\mu$.

Notice that $e_\omega^\mu$ depends on the frequency since the range of
integration is proportional to $\omega^{1/3}$. However, recall that we are
interested in high harmonics and so $\omega L\gg1$. In this limit, the term
$e_\omega^\mu$ approaches a frequency-independent form
\begin{eqnarray}
 e_\omega^\mu \rightarrow e^\mu \equiv
\frac{\Gamma(\frac{1}{3})\Gamma(\frac{2}{3})}{3L\alpha_+\alpha_-}
\left(\frac{f''^\mu_+}{\alpha_+}-\frac{f''^\mu_-}{\alpha_-}\right)~.
\end{eqnarray}

Then, Eq.~(\ref{eq:solution}) gives
\begin{eqnarray}
 A_\omega^\mu(\vec{x}) \propto e^\mu~,
\end{eqnarray}
The corresponding electric and magnetic fields are
\begin{eqnarray}
 \vec{E}_\omega(\vec{x}) &\propto& \vec{e},
\nonumber \\
 \vec{B}_\omega(\vec{x}) &\propto&
\vec{e}\times\hat{k}~,
\end{eqnarray}
where ${\hat k}$ is the unit vector in the direction of the beam
emitted from the cusp. We have used ${\vec e} \cdot {\hat k} =0$
because ${\vec f}'_+ = - {\vec f}'_- = {\hat k}$ at a cusp
and ${\vec f}''_\pm \cdot {\vec f}'_\pm =0$ because
$|{\vec f}'_\pm |=1$ [see Eq.~(\ref{stringmotion})].

The form of the electric field shows that the radiation from
cusps is linearly polarized in the direction ${\vec e}$.
Furthermore, the direction of linear polarization is independent
of the frequency of observation.

The above analysis applies for radiation exactly along the direction
of the beam. Slightly off the direction of the beam, the saddle
point in the integrals of Eq.~(\ref{eq:Jk}) will acquire small
imaginary components, and this causes the beam to die off exponentially
fast outside an angle \cite{Vilenkin:1986zz, BlancoPillado:2000xy}
\begin{equation}
\theta_\omega \simeq (\omega L)^{-1/3} ~ .
\label{beamwidth}
\end{equation}
Therefore, the width of the beam is given by $\theta_\omega$.
Similarly the beam at frequency $\omega$ is emitted for a
duration given by
\begin{equation}
 \delta t_\omega \simeq \frac{L^{2/3}}{\omega^{1/3}}~,
\label{beamduration}
\end{equation}
This is not the {\it observed} duration of the beam which we will
discuss in the next section.

Within the beam, the energy radiated in a burst per unit frequency
per unit solid angle is
\begin{eqnarray}\label{power}
\frac{d^2E_{rad}}{d\omega d\Omega} \sim
2I^2 L^2 |\vec{e}|^2 ~, \ \theta < \theta_\omega ~ ,
\end{eqnarray}
where $\theta$ denotes the angle measured from the direction of the
beam. The energy arriving at a distance $r$ is given by
\begin{eqnarray}
\frac{1}{r^2}\frac{d^2E_{rad}}{d\omega d\Omega} \sim
2\frac{I^2 L^2}{r^2} |\vec{e}|^2 ~, \ \theta < \theta_\omega ~ .
\label{Eemission}
\end{eqnarray}

Cosmic string loops are large objects and the fundamental frequency
of radiation, given by $\sim L^{-1}$, is very small. Hence, radiation
that can be observed is due to emission at very high harmonics.
Although the energy per solid angle does not depend on the frequency,
the width of the beam $\theta_\omega$ does become smaller with
increasing frequency. This suggests that the event rate will be
largest at lower frequencies where the beam is wider. Hence, it
seems favorable to seek bursts from strings in the radio band,
though the dependence of the event rate on frequency can be
more complicated because the more numerous small loops produce
higher frequencies.

We now examine the event rate in more detail.

\section{Burst event rate}
\label{sec:eventrate}

Arguments of scale invariance and simulations of a cosmic string
network indicate that the loop distribution function in the
radiation-dominated epoch is
\begin{equation}
dn_{L_0} \sim \kappa \frac{dL_0}{L_0^{5/2} t^{3/2}},
\end{equation}
where $n_{L}$ is the number density of loops of size $L$ at cosmic
time $t < t_{\rm eq}$, where $t_{\rm eq}$ is the time of
radiation-matter equality, and $\kappa \sim 1$.
In the matter-dominated epoch, $t > t_{\rm eq}$, there will
be two components to the loop distribution. The first is
the loops that were produced in the radiation-dominated
era but survived into the matter era. The second is the
loops that were produced during the matter-dominated era
and these are expected to have a $1/L^2$ distribution.
The total loop distribution is a sum of these two components,
\begin{equation}
dn_{L_0} \sim
\left ( \kappa_M + \kappa_R \sqrt{\frac{t_{\rm eq}}{L_0}} \right )
            \frac{dL_0}{L_0^{2} t^{2}},
\end{equation}
where $\kappa_R$ and $\kappa_M$ are order 1 coefficients
relevant for the radiation and matter era distributions.
We will take $\kappa_M \sim \kappa_R \equiv \kappa$.

Radiative losses from loops imply that
the loops shrink with time and so
\begin{equation}
L(t) = L_0 - \Gamma (t-t_i),
\end{equation}
where $\Gamma$ is a parameter and we will use $t \gg t_i$, i.e.,
we consider a time much later than the time when the loop was
produced. For shrinkage due to gravitational radiation,
$\Gamma \sim 100 G\mu$, where $\mu$ is the string tension, e.g., for
strings produced at the scale of $10^{14}$ GeV,
$G\mu \approx 10^{-10}$. Therefore, the loop distribution
function, taking energy losses into account, is
\begin{equation}
dn(L,t) = \frac{\kappa C_L dL}{t^2 (L+\Gamma t)^2},
\end{equation}
where
\begin{equation}
C_L \equiv  1 + \sqrt{\frac{t_{\rm eq}}{L+\Gamma t}},
\end{equation}
For $L \ll \Gamma t_0$, the radiation era loops are more important
because we will be interested in $\Gamma < 10^{-6}$ whereas
$t_{\rm eq}/t_0 \approx 10^{-5}$. For larger $L$, the
matter era loops dominate.

We now write this formula in terms of the redshift, $z$, in the matter
dominated era
\begin{eqnarray}
 dn(L,z) \simeq
      \frac{\kappa C_L (1+z)^6 dL}{t_0^2[(1+z)^{3/2} L+\Gamma t_0]^2}~,
\end{eqnarray}
where
\begin{equation}
1+z = \left ( \frac{t_0}{t} \right )^{2/3},
\end{equation}
and
\begin{equation}
C_L = 1 + (1+z)^{3/4}
            \sqrt{\frac{t_{\rm eq}}{(1+z)^{3/2} L+\Gamma t_0}}.
\label{eq:CLz}
\end{equation}
The current age of the Universe is $t_0 \simeq 4\times 10^{17}$ s.

If a loop has a cusp, there will be a burst in every period of oscillation.
So the rate of cusps on a loop of length $L$ is $c/L$ where $c \sim 1$
is the probability that a loop will contain a cusp \cite{Copi:2010jw}.
If the loop is at cosmological redshift, the observed rate of cusps
on a given loop will be $c/(L(1+z))$ due to time dilation.

The radiation from a cusp can be emitted in any direction. Only the
bursts pointing in the direction of the observer are relevant. Since
the beam width at frequency $\omega$ is
$\theta_\omega \sim (\omega L)^{-1/3}$ [Eq.~(\ref{beamwidth})],
the event rate will be suppressed by a factor $\theta_\omega^2$.

Combining all these factors gives an event ``production'' rate
in a spatial volume $dV$,
\begin{eqnarray}\label{ddotN}
 d\dot{N} \simeq
 c \frac{\theta_\omega^2}{L (1+z)} dn(L,z)dV ~,
\end{eqnarray}
Note that the beam of radiation emitted from a cusp is wider at 
lower frequencies. Thus, if a burst is observed at a particular 
frequency $\omega_e$, it will also be observed at all lower 
frequencies.

The volume element is converted to a redshift element using the
distance-redshift relation assuming a matter-dominated, flat
cosmology
\begin{equation}
H_0 dr = \frac{dz}{(1+z)^{3/2}},
\label{eq:distance}
\end{equation}
where $H_0=2/(3t_0)=72~{\rm km/sec/Mpc}$. Then,
\begin{equation}
r = \frac{2}{H_0} \left [ 1 - \frac{1}{\sqrt{1+z}} \right ] ~,
\label{eq:rz}
\end{equation}
and the physical volume is
\begin{equation}
dV = \frac{16\pi}{H_0^3} \left [ 1-\frac{1}{\sqrt{1+z}} \right ]^2
                 \frac{dz}{(1+z)^{9/2}},
\end{equation}
where we have integrated over the angular coordinates.

As a consequence, the burst production rate is
\begin{eqnarray}
d\dot{N} \simeq
\frac{A t_0 \nu_{e} C_L}{(\nu_e L)^{5/3}}
\frac{(1+z)^{-1/2} [\sqrt{1+z}-1]^2}
     {[(1+z)^{3/2}L+\Gamma t_0]^2} dL dz ,
\label{eq:prodrate}
\end{eqnarray}
where all the numerical factors have been consolidated in $A \sim 50$,
and the subscript ``e'' on $\nu_e$ denotes that it is the frequency
at emission.

From an observer's point of view, the burst production rate is not
relevant; instead, we must find the event rate that observers can
expect to see within the parameters of their instruments. Thus, the
event rate must be expressed in terms of quantities such as
the energy flux per frequency interval, $S$, to which the instrument 
is sensitive,
and the burst duration, $\Delta$, that can be detected. So, we must
transform the variables $(L,z)$ occuring in
Eq.~(\ref{eq:prodrate}) to $(S,\Delta)$. That will give the
event rate in terms of variables that are relevant to observation.

The observed frequency is related to the emitted frequency by a
redshift factor
\begin{equation}
\label{omega_o}\label{eq:nu_o}
\nu_o = \frac{\nu_e}{1+z}~.
\end{equation}
The energy flux per frequency interval can be found from
the radiated energy in Eq.~(\ref{Eemission}), which gives the
total energy radiated from the cusp. To get the energy radiated
per unit time, we need to divide that expression by the observed
duration of the burst, $\Delta$.
So the observed energy per unit time per unit area per unit 
frequency interval is
\begin{equation}
S \simeq \frac{I^2L^2}{r^2 \Delta},
\label{eq:S}
\end{equation}
where $r$ is given in terms of $z$ in Eq.~(\ref{eq:rz}).
We have chosen to normalize the cosmological scale factor
to be one today, $a(t_0)=1$, and hence the coordinate distance
$r$ is also the physical distance at the present epoch. As a simplifying assumption we will only consider the case 
when the current, $I$, is a constant. In general, the current will 
depend on the cosmological epoch because it can build up due 
to string interactions with a cosmological magnetic field
and dissipate due to scattering of the charge carriers
on the string.

The duration of the burst is determined by a combination of the
duration at emission (``intrinsic'' duration), the cosmological
redshift, and the time delays due to scattering with the cosmological
medium. This last factor is important for bursts at long wavelengths
such as in the radio. The intrinsic burst duration at the emission
point is given in Eq.~(\ref{beamduration}). To obtain the 
burst duration at the observation point, we have to include a factor 
of $1/\gamma^2$ where $\gamma \sim (\omega L)^{1/3}$ is the Lorentz
factor at the cusp. This factor was first derived in Ref.~\cite{Babuletal} 
and is also seen in synchrotron radiation \cite{RybickiLightman}; it 
was, however, missed in Ref.~\cite{Vachaspati:2008su}. It arises
because the cusp is moving toward the observer and so photons
emitted over a time interval $\delta t$ arrive at the observer
in the interval $\delta t (1-v)$ where $v$ is the speed of the
string in the emitting region near the cusp. In addition, we need 
to include 
a cosmological redshift factor to account for time dilation. This
gives the intrinsic beam duration at the observation point 
\begin{equation}
\label{Delta_inS}
 \Delta{t}_{in} \simeq 
 \frac{(1+z)L^{2/3}}{\nu_e^{1/3}}\frac{1}{(\nu_e L)^{2/3}}
                \simeq \frac{1}{\nu_o}~.
\end{equation}

The burst duration due to scattering with the turbulent intergalactic
medium at given frequency, $\nu_o$, and redshift, $z$, is modeled
as a power law
\cite{LeeJokipii1976,Kulkarnietal}
(for a review, see \cite{Rickett:1977vv})
\begin{equation}
 \Delta{t}_{s} \simeq \delta{t}_1 \left (\frac{1+z}{1+z_1}\right )^{1-\beta}
                   \left ( \frac{\nu_o}{\nu_1} \right )^{-\beta} ~,
\label{eq:deltatS}
\end{equation}
where, empirically,
\begin{equation}\label{paraexp}
\delta t_1 = 5~{\rm ms}, \  z_1 = 0.3, \ \nu_1 = 1.374 ~ {\rm GHz},
\ \beta = +4.8.
\label{emperical}
\end{equation}
Note that with our conventions in Eq.~(\ref{eq:deltatS}), $\beta > 0$.
(In Ref.~\cite{Vachaspati:2008su} the sign conventions were such that
$\beta$ was negative.)

The total burst duration, $\Delta$, is a sum in quadratures of
the intrinsic time width and the width due to scattering
\begin{equation}
\label{Delta}
 \Delta = \sqrt{\Delta{t}_{in}^2+\Delta{t}_{s}^2}~,
\label{eq:Delta}
\end{equation}
Inserting the expressions in Eqs.~(\ref{Delta_inS}) and 
(\ref{eq:deltatS}) leads to
\begin{equation}
1+z = \frac{(\Delta^2 \nu_o^2 -1)^{1/2(1-\beta)}}
           {\delta_1^{1/(1-\beta)} \nu_o} \ ,
\label{eq:1+z}
\end{equation}
where
\begin{equation}
\delta_1 \equiv \nu_1^\beta \delta t_1 (1+z_1)^{\beta-1} \ .
\end{equation}
Inserting numerical values from Eq.~(\ref{emperical}) gives
\begin{equation}
1+z \simeq \frac{82}{(\Delta^2\nu_o^2-1)^{1/2(\beta-1)}} 
          \left ( \frac{\nu_1}{\nu_o} \right ) \ .
\end{equation}
Our calculations assume that $z < z_{\rm rec}\simeq 1100$, the 
redshift at recombination. Then, the constraints $0<z<z_{\rm rec}$
give
\begin{equation}
\Delta_{\rm min} < \Delta < \Delta_{\rm max} \ ,
\end{equation}
where
\begin{equation}
\Delta_{\rm min} = \frac{1}{\nu_o} \left \{
 1 + \left [ 0.075 \left ( \frac{\nu_1}{\nu_o} \right ) \right ]^{2(\beta-1)}
      \right \}^{1/2} \ ,
\label{Deltamin}
\end{equation}
\begin{equation}
\Delta_{\rm max} = \frac{1}{\nu_o} \left \{
 1 + \left [ 82 \left ( \frac{\nu_1}{\nu_o} \right ) \right ]^{2(\beta-1)}
      \right \}^{1/2} \ .
\label{Deltamax}
\end{equation}
For example, with $\nu_o=\nu_1$, 
$\Delta_{\rm min} \sim 7\times 10^{-10}~{\rm s}$
and $\Delta_{\rm max} \sim 1.3\times 10^{-2}~{\rm s}$.

To transform from intrinsic variables $(L,z)$ to observer
variables $(S,\Delta)$, we need to calculate the Jacobian of
the transformation. We have already obtained $z(\Delta, S)$
in Eq.~(\ref{eq:1+z}). From Eq.~(\ref{eq:S}), we also obtain 
\begin{equation}
\label{eq:L}
 L = \frac{r}{I} \sqrt{S \Delta},
\end{equation}
and $r$ is a function of $z$ [Eq.~(\ref{eq:rz})] which is a function
of $(S,\Delta)$ as in (\ref{eq:1+z}).
Some algebra then leads to the Jacobian factor
\begin{equation}
\left | \frac{\partial (L, z)}{\partial (S,\Delta)} \right | = 
\frac{\nu_o L\Delta }{2 (\beta-1)\delta_1^{1/(1-\beta)} 
       S (\Delta^2\nu_o^2 -1)^{1-1/2(1-\beta)}} \ ,
\end{equation}
where $L=L(S,\Delta)$ via Eq.~(\ref{eq:L}).

Now, we can get the event rate in observer variables from the
production rate of Eq.~(\ref{eq:prodrate}),
\begin{eqnarray}
d\dot{N} &\simeq& \frac{A t_0}{2(\beta-1)} 
\frac{C_L \nu_{o}^2 \Delta}{S (\nu_o L)^{2/3}}
 \frac{1}{\Delta^2 \nu_o^2 -1}
        \nonumber \\
     && 
     \times \frac{[\sqrt{1+z}-1]^2}
     {(1+z)^{1/6}[(1+z)^{3/2}L+\Gamma t_0]^2} dS d\Delta \ ,
\label{ddotNfinal}
\end{eqnarray}
where we have used Eq.~(\ref{eq:nu_o}), $L$ is given by Eq.~(\ref{eq:L}),
and $r(z)$ by Eq.~(\ref{eq:rz}). Note that $\Delta_{\rm min}^2 \nu_o^2 > 1$ 
and so the event rate does not have a singularity for 
$\Delta \in [\Delta_{\rm min}, \Delta_{\rm max}]$.

Equation (\ref{ddotNfinal}) is our final expression for the differential
event rate. We will now analyze the expression to extract certain
closed form results.

\section{Event Rate Analysis}
\label{sec:analysis}

Even though the emitted burst duration in Eq.~(\ref{beamduration})
depends on the length of the
loop, the observed burst duration in Eq.~(\ref{eq:Delta}) is independent 
of the length of the loop. Hence, for a given burst duration at a 
certain frequency, a loop (of any length) has to be at the redshift
given by Eq.~(\ref{eq:1+z}). This fixes the distance to the loop.
The energy flux from a loop, however, does depend on its length and
Eq.~(\ref{eq:L}) gives $L \propto \sqrt{S}$. The only implicit dependence
of the event rate in Eq.~(\ref{ddotNfinal}) on $S$ occurs through
$L$ which also appears in $C_L$. If we consider the limit 
$S \to \infty$, we have $(1+z)^{3/2} L \gg \Gamma t_0$ and $C_L \to 1$,
and simple power counting gives
\begin{equation}
d{\dot N} \propto \frac{dS}{S^{7/3}} \ , \ \ S \to \infty \ .
\label{dNdothighS}
\end{equation}
For smaller $S$, such that $(1+z)^{3/2}L \ll \Gamma t_0$, the
power counting gives
\begin{equation}
d{\dot N} \propto \frac{dS}{S^{4/3}} \ . 
\label{dNdotlowS}
\end{equation}
   
The dependence of the event rate on the burst duration is less
apparent. Note that long duration bursts, $\Delta \gg 1/\nu_o$,
are only possible if the loop is very close, and then too it is
not possible to have bursts of arbitrarily long duration at some
fixed observation frequency. The maximum possible duration occurs
at $z=0$ and is given in Eq.~(\ref{Deltamax}). So, to find the event 
rate for duration bursts close to $\Delta_{\rm max}$, we expand the 
event rate around $z=0$. First, we obtain
\begin{equation}
z \simeq \frac{\nu_o^{2\beta} \Delta_{\rm max}}{(\beta-1)\delta_1^2}
           (\Delta_{\rm max}-\Delta) \ ,
\end{equation}
which leads to
\begin{equation}
d{\dot N} \propto 
  \left ( \frac{\Delta_{\rm max}-\Delta}{S} \right )^{4/3} dS d\Delta \ ,
\end{equation}
where we have assumed $L \ll \Gamma t_0$ which implies that $S$ cannot
be too large.

Having obtained these limiting forms for the event rate, we now 
turn to a numerical evaluation.

\section{Numerical Estimates}
\label{sec:numerical}

We now find the event rate as a function of the flux and the burst 
duration by numerically evaluating and integrating Eq.~(\ref{ddotNfinal}).

For our numerical estimates, we take the cosmological parameters
\begin{equation}
t_0 = 4 \times 10^{17} ~{\rm s} \ , \ \
t_{\rm eq} = 2.4 \times 10^{12} ~{\rm s} \ .
\end{equation}
We also assume the string parameters
\begin{equation}
I = 10^5 ~{\rm GeV} \ , \ \
\Gamma = 10^{-8}.
\end{equation}
Typically, for string loop decay due to gravitational radiation,
$\Gamma \sim 100 G\mu$ where $G$ is the gravitational Newton's
constant and $\mu$ is the string tension. Therefore, our choice of
$\Gamma$ corresponds to $G\mu \sim 10^{-10}$ or a symmetry breaking
energy scale of $10^{14}~{\rm GeV}$, which is a scale at which
Grand Unification may occur.

The scattering contribution to the burst duration in
Eq.~(\ref{eq:deltatS}) contains a number of parameters that are determined
empirically, and are shown in Eq.~(\ref{paraexp}). In exploring parameter
space, we shall assume a range of parameters motivated by the Parkes
survey \cite{Lorimer:2007qn},
\begin{eqnarray}
\nu_o &\in& (1.230,1.518) ~{\rm GHz}\ , \nonumber \\
\Delta &\in& (10^{-3},1) ~{\rm s} \ , \nonumber \\
S &\in& (10^{-5},10^{+5}) ~{\rm Jy} \ .
\label{eq:ranges}
\end{eqnarray}
Note the conversion
\begin{equation}
1~{\rm Jy} = 10^{-23} \frac{\rm ergs}{\rm cm^2-s-Hz}.
\end{equation}

%With the above parameters, $a$ and $b$ can be written as
%\begin{equation}
%a = 2.58 \times 10^{-4}\, \nu_{o, {\rm GHz}}^{9.6}\, \Delta_{\rm ms}^{2},
%\end{equation}
%\begin{equation}
%b = 1.88 \times 10^{-4}\, I_{5}^{-4/3}\, \nu_{o, {\rm GHz}}^{8.93}\,
%\Delta_{\rm ms}^{2/3}\, S_{\rm Jy}^{2/3},
%\end{equation}
%where $\nu_{o,{\rm GHz}} \equiv \nu_{o}/1\,$GHz,
%$\Delta_{\rm ms} \equiv \Delta/ 1\,$ms,
%$I_{5} \equiv I/ 10^{5}\,$GeV and
%$S_{\rm Jy} \equiv S/1\,$Jy.
%Note the conversion
%\begin{equation}
%1~{\rm Jy} = 10^{-23} \frac{\rm ergs}{\rm cm^2-s-Hz}.
%\end{equation}

\begin{figure*}[htbp]
\includegraphics[scale=0.3]{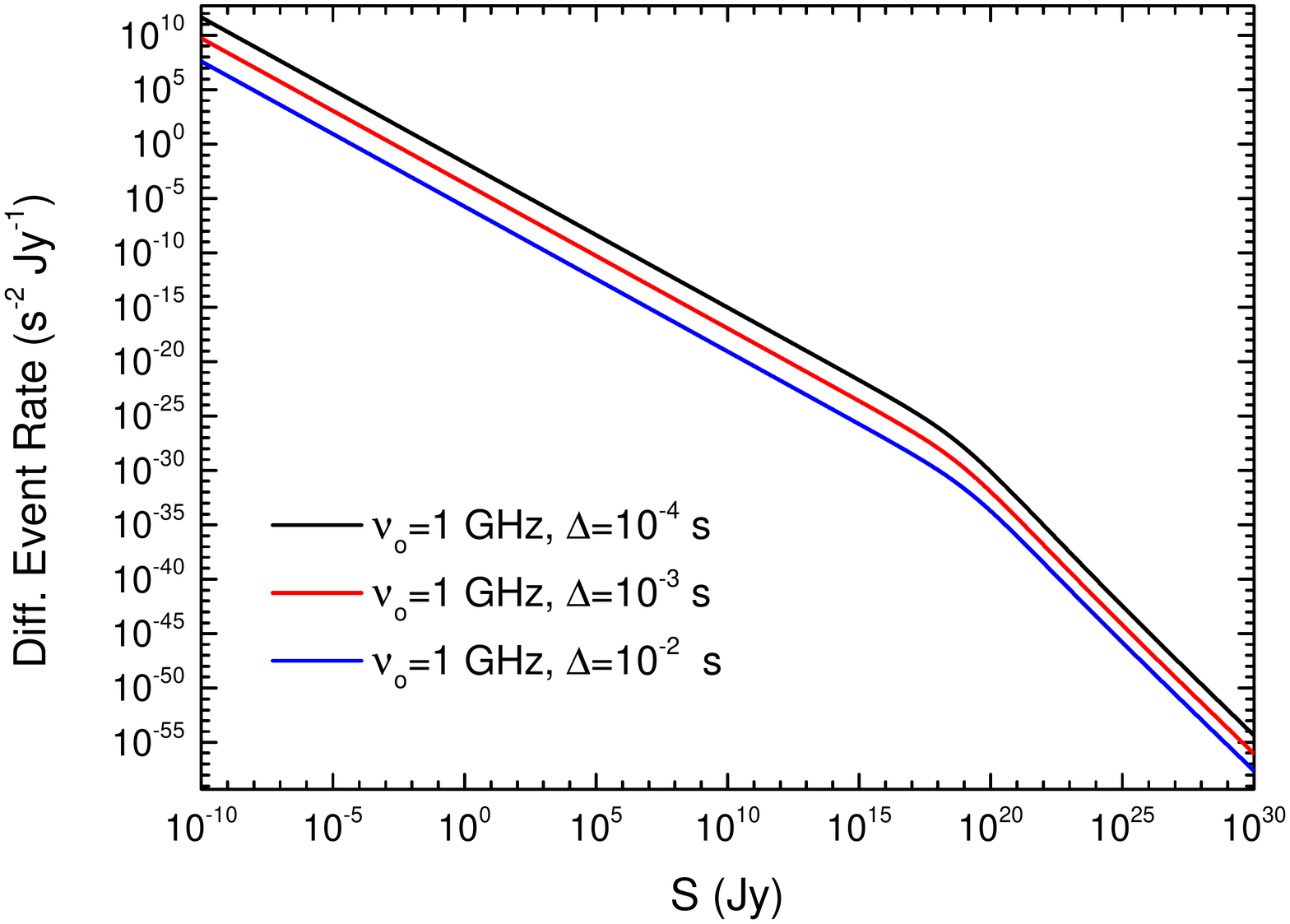}
\includegraphics[scale=0.3]{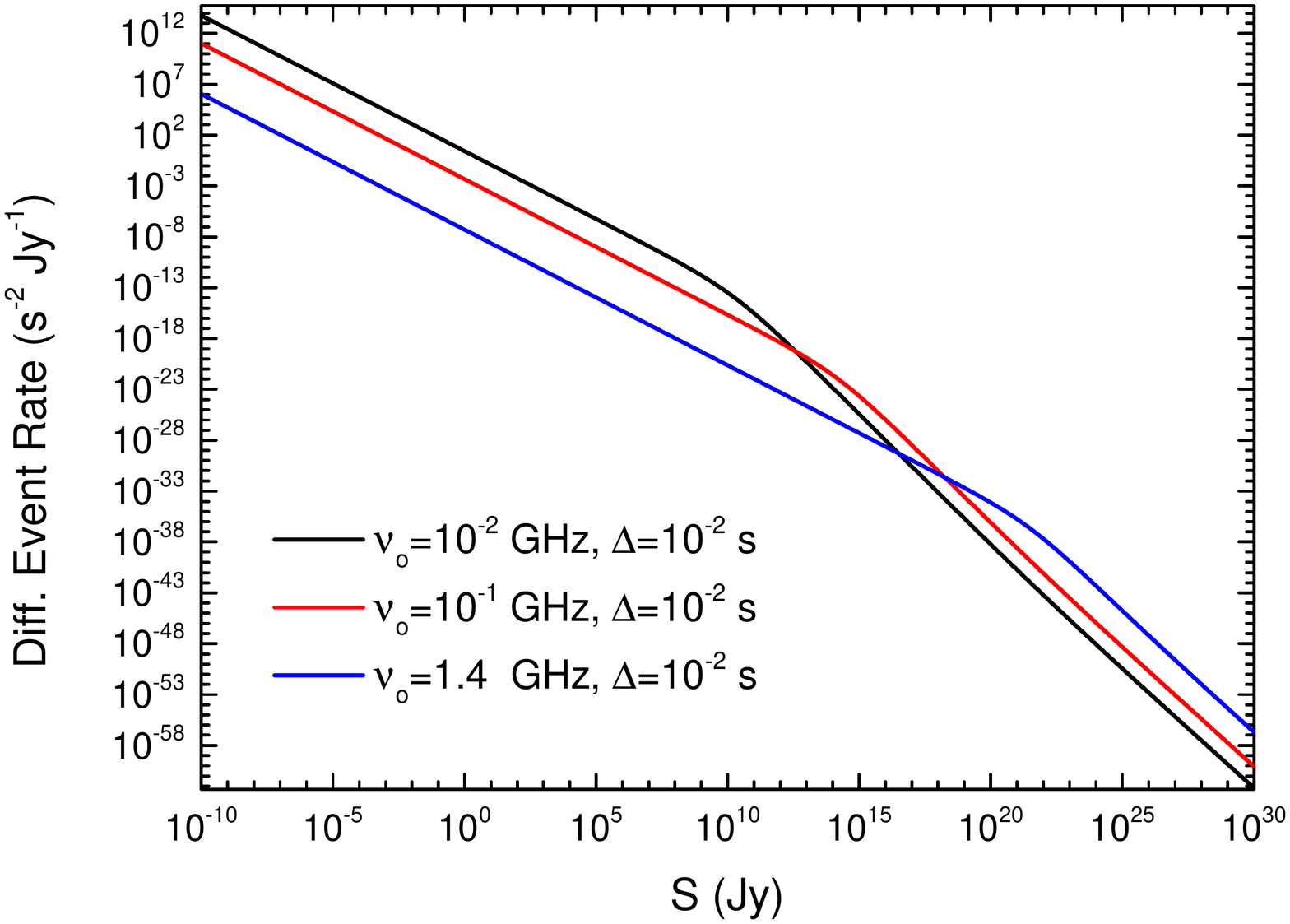}
\caption{
The differential event rate of radio bursts emitted from
superconducting cosmic strings as a function of flux $S$. In the
left panel, the observed frequency is fixed, $\nu_o=1 ~{\rm GHz}$,
and the duration is chosen to be 
$\Delta=10^{-4}, ~10^{-3}, ~10^{-2}~{\rm s}$
(top to bottom curves). In the right panel, the duration is fixed,
$\Delta=0.01 ~{\rm s}$, and the observed frequency is chosen to be
$\nu_o=0.01, ~0.1, ~1.4 ~{\rm GHz}$ (top to bottom curves for small $S$).}
\label{Fig:Event_a}
\end{figure*}

In the following figures, we show the differential event rate of 
Eq.~(\ref{ddotNfinal}) as functions of the flux and burst duration. 
First, in Fig.~\ref{Fig:Event_a} we plot the differential event rate 
as a function of $S$ for several different choices of $\nu_o$ and 
$\Delta$. The plot is
made on a log-log scale to accommodate the wide range of scales, and 
shows two different power law behaviors, consistent with the
analytical results in Eqs.~(\ref{dNdothighS}) and (\ref{dNdotlowS}). 

%This divergence is the sweet spot. Also, we have cut-off the plot if the
%value of the redshift determined from 
%Eq.~(\ref{determine_z}) exceeds
%the redshift at recombination, $z_{\rm rec}=1100$. For example, the plot
%with $\nu_o=1.4~{\rm GHz}$ and $\Delta=0.1~{\rm s}$ is cut-off by the
%dashed line for $S \sim 10^{-3} ~{\rm Jy}$ because smaller values of $S$
%could only come from loops at redshifts larger than $z_{\rm rec}$ and
%the different cosmology ({\it e.g.} ionized medium) at those epochs is not
%taken into account in our calculation.

\begin{figure*}[htbp]
\includegraphics[scale=0.3]{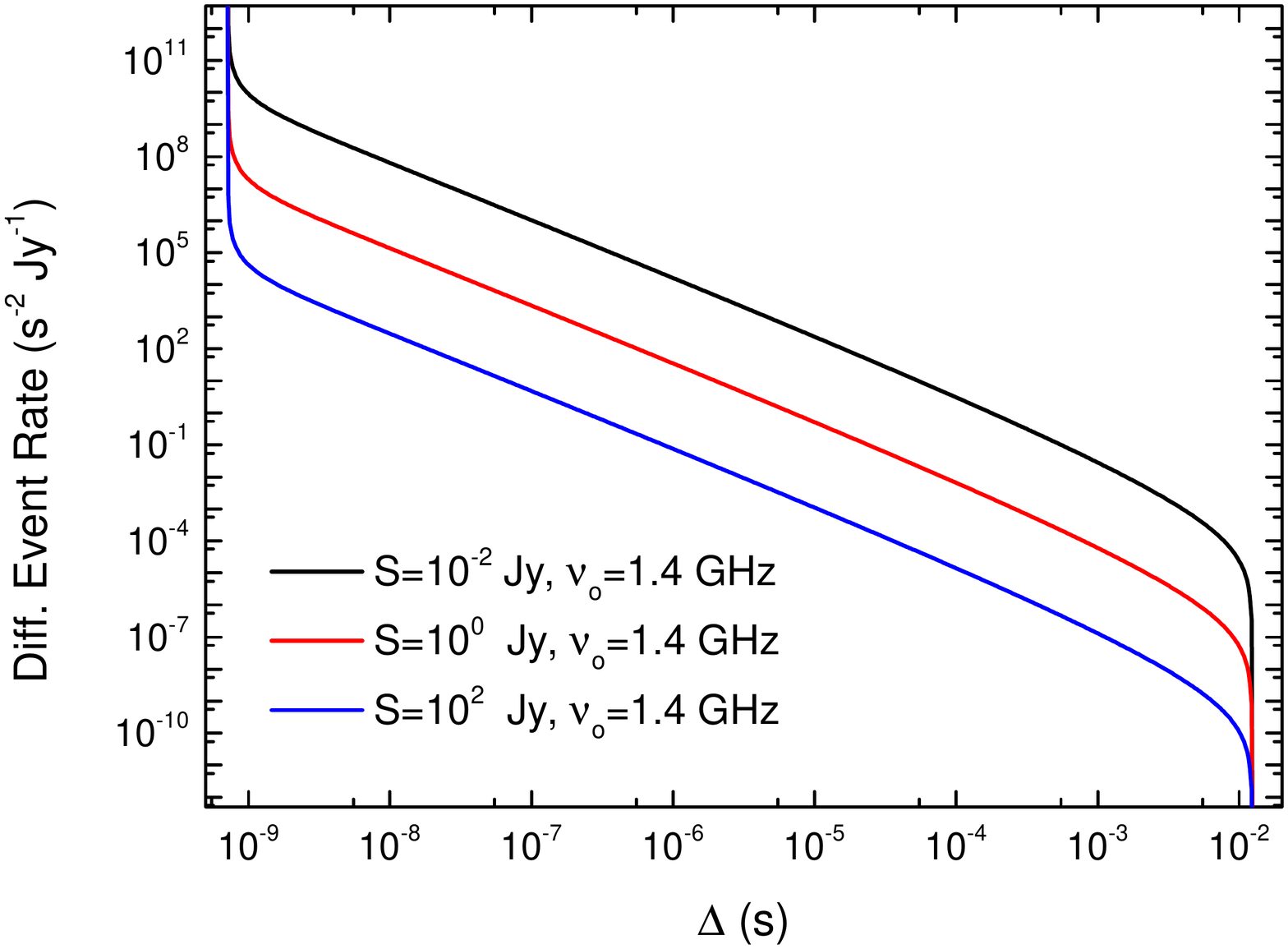}
\includegraphics[scale=0.3]{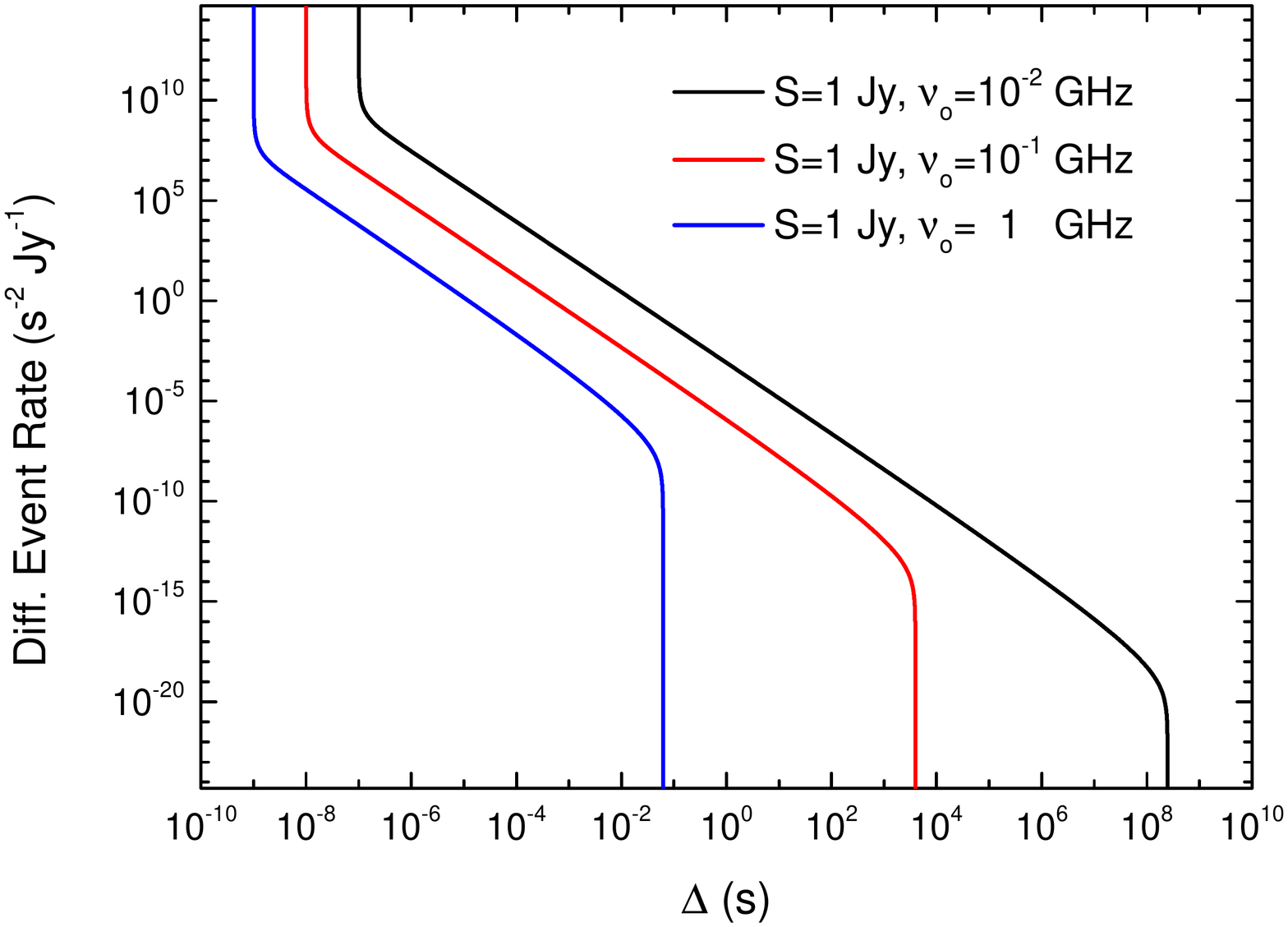}
\caption{
The differential event rate of radio bursts from superconducting
cosmic strings as a function of duration $\Delta$. In the left panel, the
observed frequency is fixed, $\nu_o=1.4 ~{\rm GHz}$, and the flux is
chosen to be $S=10^{-2}, ~1, ~10^2 ~{\rm Jy}$ (top to bottom curves).
In the right panel, the flux is fixed as $S=1 ~{\rm Jy}$, and the observed
frequency is chosen to be $\nu_o=0.01, ~0.1, ~1 ~{\rm GHz}$ (top to bottom
curves). The range of $\Delta$ lies in $[\Delta_{\rm min},\Delta_{\rm max}]$
as defined in Eqs.~(\ref{Deltamin}), (\ref{Deltamax}).}
\label{Fig:Event_c}
\end{figure*}

In Fig.~\ref{Fig:Event_c}, we show the dependence of the differential 
event rate on the burst duration for a variety of values of $S$ and 
$\nu_o$. 

\begin{figure*}
\includegraphics[scale=0.3]{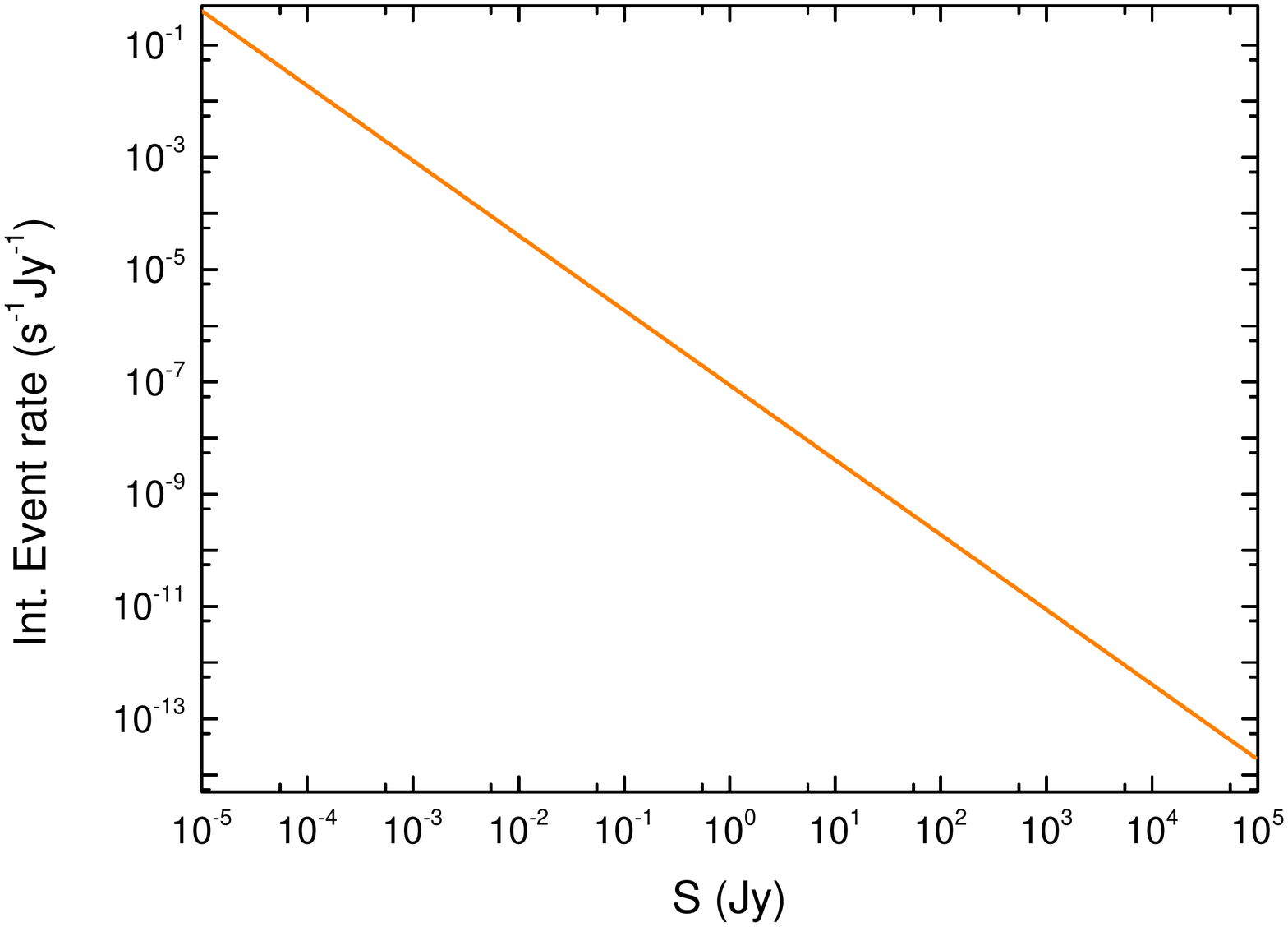}
\includegraphics[scale=0.3]{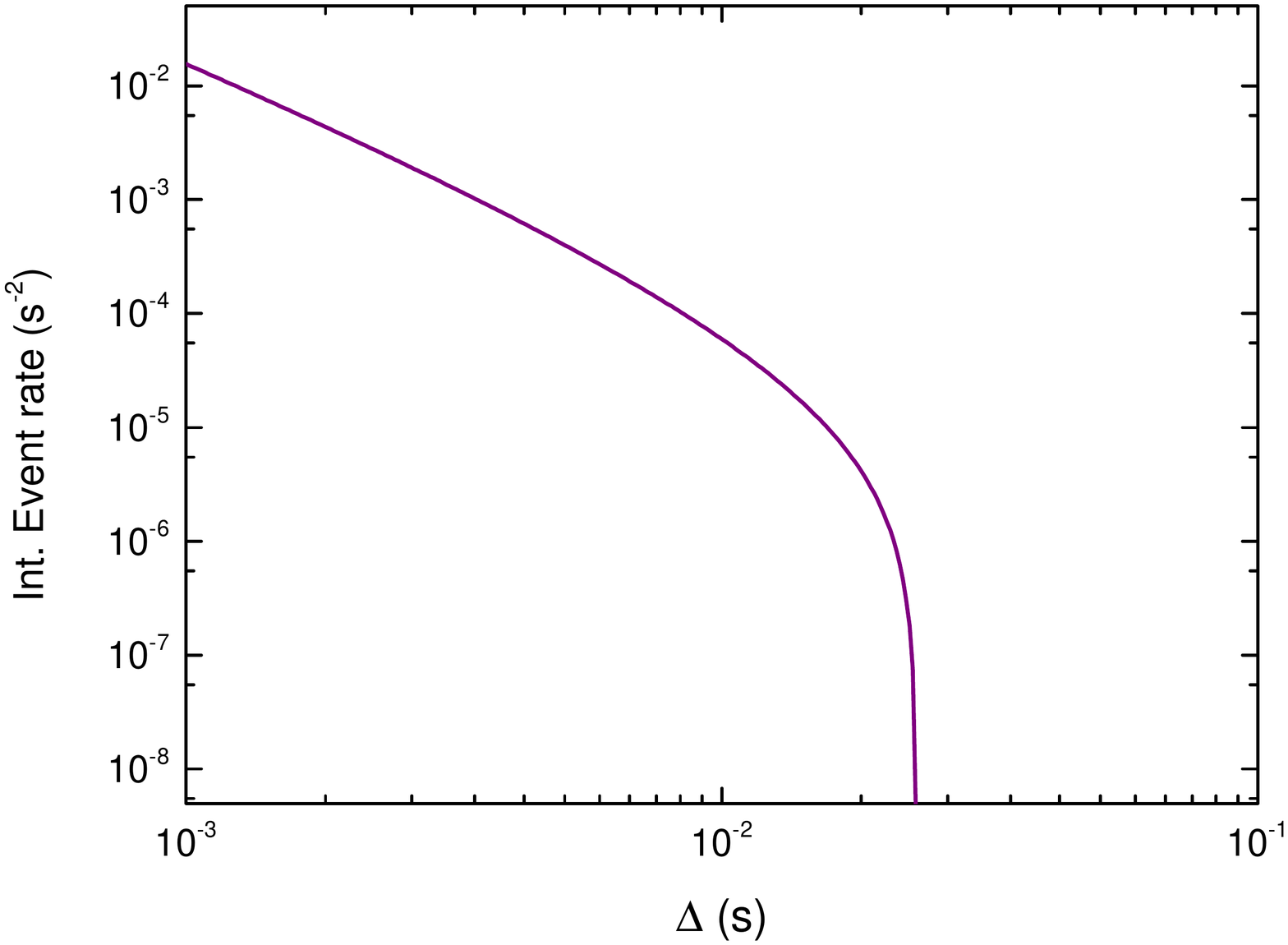}
\caption{The integrated event rate of radio bursts with fixed 
observable frequency, $\nu_o=1.23~{\rm GHz}$, from superconducting 
cosmic strings as functions of the flux $S$ (left panel) and the 
duration $\Delta$ (right panel). The intervals of integration and 
other parameters are given in Sec.~\ref{sec:numerical}.}
\label{Fig:Event_in}
\end{figure*}

The integrated event rate as a function of the flux $S$ and burst duration
$\Delta$ is shown in the left and right panels of Fig.~\ref{Fig:Event_in}.
The asymptotic fits to these plots are
\begin{eqnarray}
\frac{d\dot{N}}{dS} &\simeq&  
10^{-7} \left(\frac{S}{1~ {\rm Jy}}\right)^{-4/3} 
                                 {\rm s}^{-1} {\rm Jy}^{-1} ~, \\
\frac{d\dot{N}}{d\Delta} &\simeq&  10^{-2} 
   \left(\frac{\Delta}{1~ {\rm ms}}\right)^{-9/4} {\rm s}^{-2} ~.
\end{eqnarray}
Hence, an experiment that integrates events over the ranges 
of $\Delta$ in Eq.~(\ref{eq:ranges}), and is sensitive to milli Jansky 
fluxes, will observe on the order of $1$ radio bursts per month, if 
there are superconducting cosmic strings with the chosen parameters. 
Turning this figure around, a search for cosmological radio transients 
can place stringent constraints on superconducting cosmic strings. 
If we consider radio bursts emitted by superconducting strings with 
observable frequency 1.23 {\rm GHz} and flux greater than $300~{\rm mJy}$, 
the event rate is $\sim 10^{-3}$ per hour, and is a factor of 10 smaller 
than the upper bound given by the Parkes survey \cite{Lorimer:2007qn}, 
$0.025$ per hour. Since the predicted event rate depends on
the string parameters, this result implies that current radio experiments 
already rule out an interesting part of parameter space (current and 
string tension).

\section{Conclusions}
\label{conclusions}

We have addressed two observational aspects of radio transients produced
by cusps on superconducting strings. First, we have shown that the radiation
emitted along the beam direction is linearly polarized, and the direction
of polarization is independent of the frequency. The polarization
can be used as a discriminating signature for radio bursts from
superconducting strings, though a more detailed study should
also consider the dependence of the polarization as a function of angle
from the direction of the beam and the variation in the polarization
over the duration of the event. Second, we have calculated the event
rate of radio bursts from cusps on superconducting strings in terms of
observational variables, namely, the burst duration and the flux. Our 
calculation includes the Jacobian that results from the transformation 
from string variables to observational variables.

Unlike burst events in higher energy parts of the electromagnetic
spectrum, a novelty of the calculation for radio bursts is that the 
burst duration depends on the redshift of the burst event due to 
two contributions: the cosmological redshift and the scattering due 
to intervening matter. As is well understood,
the former grows with redshift as $1+z$ and when the redshifting
of the frequency is also taken into account, gives a duration of
$1/\nu_o$. The contribution of scattering is
given by Eq.~(\ref{eq:deltatS}) \cite{LeeJokipii1976,Kulkarnietal}
and is somewhat counterintuitive because it diminishes with increasing
redshift. To understand this (partially), we note that the burst duration 
increases due to scattering because scattering allows photons to bend 
into the direction of the observer. If the relevant scattering can be 
thought to occur at roughly half the distance to the source,
for a source that is farther away, the halfway scattering point
is also more distant. Therefore, for fixed observational frequency,
the frequency at the scattering point is also higher, and hence
scattering is less efficient. Thus more distant bursts get a smaller
contribution to their duration from the scattering. The two
contributions to the burst duration are added in quadrature,
yielding Eq.~(\ref{eq:Delta}).

%The addition of two components in the burst duration, one that
%grows with redshift and the other that decreases with redshift
%is important for us because it implies that there is a minimum
%burst duration that occurs at a definite redshift. In other words,
%there is a redshift for which
%$\partial \Delta /\partial z$ vanishes, and this gives a
%divergent contribution in the transformation Jacobian, and hence
%a sweet spot. Furthermore, the sweet spot occurs at parameter
%values that are within the parameter range of experiments. For
%example, for $b \ll 1$ the sweet spot location is given by
%Eq.~(\ref{sslocation}), which in numerical values can be
%re-written as
%\begin{equation}
%\Delta_{\rm ms} =
%4 I_5^{-0.71} \nu_{o,{\rm GHz}}^{-1.73} S_{\rm Jy}^{0.36}.
%\end{equation}

We have also found the integrated event rate as a function
of the flux and burst duration. For the canonical set of
parameters listed in Sec.~\ref{sec:numerical}, the integrated
event rates are reasonable, at the level of one event
per month. Such event rates indicate that the 
search for radio bursts can serve as excellent probes of the
superconducting string model.

Our analysis has been performed under some simplifying
assumptions that may need to be reexamined in the future.
Our formula for the burst duration due to scattering of
radio waves, Eq.~(\ref{eq:deltatS}), should be reexamined
in the cosmological context, since the relevant cosmological
epochs are concurrent with reionization, formation of
large scale structure, and other astrophysical activity.
Note that we have also neglected the cosmological acceleration
which will dilute the number density of cosmic strings and thus
reduce the event rate of radio bursts at low redshifts.
We have also sharply cut off all radio bursts prior to the
epoch of recombination. In principle, there will be a
gradual cut-off, though this may not make much difference
to the final results. From the string side, we have assumed
a constant current on all strings, whereas we expect the
current to grow as a string cuts through ambient magnetic fields.
If a primordial magnetic field exists, our assumption may
be justified. In the absence of a primordial magnetic field,
currents on strings will build up only after structures have 
generated magnetic fields. We have also assumed that the dominant 
energy loss from strings is due to gravitational
radiation and not due to electromagnetic losses, {\it i.e.}
$\Gamma\mu =100G\mu^2 \gg 10 I\sqrt{\mu}$. For
$I \sim 10^5~{\rm GeV}$,
this is valid if the string energy scale is larger than
$10^{14}~{\rm GeV}$, {\it i.e.} $G\mu > 10^{-10}$. For yet lighter
strings, $\Gamma$ will be set by electromagnetic losses, and
for very light strings, $\mu \sim (1~{\rm TeV})^2$,
the strings are dragged by the cosmological plasma, at least on
large length scales, and the string dynamics will be very different.
In the regime where gravitational losses dominate and radio bursts
due to short loops dominate the event rate, our numerical results
give
\begin{equation}
{\dot N} \simeq 2 \times 10^{-5} \mu_{-8}^{-5/2} I_{5}^{2/3} ~{\rm{s^{-1}}} \ , \ \
100G\mu^2 > 10 I \sqrt{\mu},
\end{equation}
where $\mu_{-8} \equiv G \mu / (10^{-8})$ and $I_{5} \equiv I / (10^{5} ~ \rm{GeV})$. If the string parameters are such that the power lost to
electromagnetic radiation is larger than that to gravitational
radiation, we should replace the expression for gravitational
power emission, $100 G\mu^2$, by the electromagnetic power
$10 I \sqrt{\mu}$. This occurs when $I > 1.2 \times 10^{8} \mu_{-8}^{3/2} ~{\rm{GeV}}$. Then,
\begin{equation}
{\dot N} \simeq 2 \times 10^{-3} \mu_{-8}^{5/4} I_{8}^{-11/6} ~{\rm{s^{-1}}} \ , \ \
100G\mu^2 < 10 I \sqrt{\mu}.
\end{equation}

There are several radio telescopes currently in operation
searching for radio transients, e.g., Parkes \cite{Lorimer:2007qn}, ETA \cite{eta}, LWA \cite{lwa}, LOFAR \cite{lofar}, and others under
construction, e.g., SKA \cite{ska}. It would be useful to tailor the analysis in our
paper to the specific range of observational parameters that will be employed
in these searches.

Cosmic string cusps also produce gravitational wave
bursts \cite{Damour-Vilenkin}, which can be detectable by sensitive
interferometers such as LIGO, VIRGO, and LISA, ultra high energy neutrino
bursts \cite{BOSV}, which can be detectable by the space-based cosmic
ray detector JEM-EUSO and by radio telescopes LOFAR and SKA via Askaryan
effect \cite{Askaryan}. There has already been some initiative to look
for electromagnetic counterparts of gravitational wave bursts at LIGO
and VIRGO \cite{ligo-virgo}. Linearly polarized radio signal and
simultaneous detection of accompanying bursts from the same cusp can
help distinguish cosmic strings from astrophysical sources, and hence
help to discover cosmic strings or to put constraints on superconducting string parameters.

%\vspace{1cm}
\acknowledgments
We are especially grateful to Ken Olum for pointing out an error in
an earlier version, and to Jose Blanco-Pillado, Robert Brandenberger, 
Mike Kavic, Ben Shlaer, John Simonetti, Daniele Steer, Alex Vilenkin, and Zheng Zheng for discussions. TV is grateful to the
Institute for Advanced Study for hospitality. This work was supported
by a grant from the Department of Energy at ASU.

\bibstyle{aps}
%\bibliography{paper}

\end{document}